\documentclass[aps,prb,amsmath,amssymb,superscriptaddress,twocolumn,10pt]{revtex4-1}

\usepackage{amssymb, amsmath, esint}
\usepackage{graphicx,color}
\usepackage{graphics}
\usepackage{subfigure}
\usepackage{comment}
\usepackage{bbold}

\usepackage[colorlinks]{hyperref}
\hypersetup{
    linkcolor=blue,
}

\usepackage{epsfig}
\usepackage{dcolumn}% Align table columns on decimal point
\usepackage{bm}% bold math

\usepackage{esint}

\usepackage{scrhack}
\usepackage{physics}
\usepackage{eurosym}

\renewcommand{\ref}[1]{S\ref{#1}}

\def\s{\bm{s}}
\def\S{\bm{S}}
\def\h{\bm{h}}

\def\H{\mathcal{H}}
\def\PT{\mathcal{PT}}
\def\P{\mathcal{P}}
\def\T{\mathcal{T}}
\def\C{\mathbb{C}}

\def\ex{\mathbf{\hat x}}
\def\ey{\mathbf{\hat y}}

\DeclareMathOperator{\sinc}{sinc}

\begin{document}

\title{Parity-time symmetry breaking in spin chains}

\author{Alexey Galda}
\affiliation{James Franck Institute, University of Chicago, Chicago, Illinois 60637, USA}
\affiliation{Materials Science Division, Argonne National Laboratory, 9700 S. Cass Avenue, Argonne, Illinois 60439, USA}
\author{Valerii M. Vinokur}
\affiliation{Materials Science Division, Argonne National Laboratory, 9700 S. Cass Avenue, Argonne, Illinois 60439, USA} 
\date{\today}

\begin{abstract}
	We investigate nonequilibrium phase transitions in classical Heisenberg spin chains associated with spontaneous breaking of parity-time ($\PT$) symmetry of the system under the action of Slonczewski spin-transfer torque (STT) modeled by an applied \textit{imaginary} magnetic field. We reveal the STT-driven $\PT$ symmetry-breaking phase transition between the regimes of precessional and exponentially damped spin dynamics and show that its several properties can be derived from the distribution of zeros of the system's partition function, the approach first introduced by Yang and Lee for studying equilibrium phase transitions in Ising spin chains. The physical interpretation of imaginary magnetic field as describing the action of non-conservative forces opens the possibility of direct observations of Lee-Yang zeros in nonequilibrium physical systems.
\end{abstract}

\maketitle

Phase transitions where physical systems experience nonanalytic changes of their properties are one of the most remarkable phenomena occurring in many-particle systems~[\onlinecite{Callen}]. In 1952 Yang and Lee devised an approach that reveals a deep structure of singularities associated with phase transitions via investigation of points on the complex plane of physical parameters where the partition function of a system vanishes (Lee-Yang zeros)~[\onlinecite{Yang:1952,Lee:1952}]. Recently, the Lee-Yang description was generalized to nonequilibrium systems~[\onlinecite{Blithe:2002,Bena:2005,Polkovnikov:2013,Hickey:2014,Pekola:2017}], and is now becoming a powerful tool promising to bring unified understanding of both equilibrium and nonequilibrium processes. The latter can be efficiently described in a framework of non-Hermitian Hamiltonian approach in which non-Hermiticity is proportional to the external bias~[\onlinecite{Galda16a,Tripathi:2016}]. Since the Lee-Yang approach rests on analytical continuation of the partition function to the complex plane of the controlling parameter, it opens an appealing opportunity for incorporating the Lee-Yang approach into a non-Hermitian scheme to enable a universal unified description of phase transitions in open dissipative systems. Here we meet the challenge and investigate nonequilibrium phase transitions in classical Heisenberg spin chains that offer an exemplary laboratory for both Lee-Yang model and non-Hermitian approach. We find the non-equilibrium phase transition associated with spontaneous breaking of parity-time ($\PT$) symmetry of the system under the action of Slonczewski spin-transfer torque (STT) modeled by imaginary magnetic field. We relate the singularities of this phase transition with the distribution of the Lee-Yang zeros. 
%Our findings open the route to direct observations of Lee-Yang zeros in out-of-equilibrium physical systems and pave the ground for a unified theory of nonequilibrium phase transitions in interacting many-body dissipative systems.

\bigskip

%%%%%%%%%%%%%%%%%%%%%%%%%%%%%%-INTRODUCTION-%%%%%%%%%%%%%%%%%%%%%%%%%%
The pioneering Yang-Lee description of phase transitions~[\onlinecite{Yang:1952,Lee:1952}] via the distribution of zeros of the partition function was achieved by going to the extended complex plane of the applied magnetic field upon adding its imaginary component. The zeros were located on a unit circle in the complex fugacity plane, $\xi = \exp(-H/k_\mathrm{B}T)$, where $H$ is the applied transverse external magnetic field measured in energy units. Later, the Lee-Yang circle theorem has been generalized to Heisenberg models~[\onlinecite{Suzuki68b, Katsura62, Suzuki69, Kawabata70, Trotter59, Heilmann70, Kunz70}], ferromagnetic Ising models of arbitrary high spin~[\onlinecite{Asano68a, Asano68b, Suzuki68a, Suzuki68b, Griffiths69}], Heisenberg and general Ising models with multiple-spin interactions~[\onlinecite{SuzukiFisher70}], and isotropic classical spins of arbitrary dimensionality on a one-dimensional lattice~[\onlinecite{Fisher64, Stanley69}].
Recently, the Lee-Yang zeros approach has emerged in nonequilibrium physics (see, e.g., Ref.~[\onlinecite{Pekola:2017}]), where the Lee-Yang zeros of a partition function of \textit{trajectories} describing the evolution of a stochastic process were studied experimentally in the non-equilibrium settings, where they characterize dynamic phase transitions occurring in a quantum system after a quench. In parallel, it has been discovered that the action of some non-conservative forces, such as Gilbert damping and Slonczewski STT, on individual spins is equivalent to switching on an imaginary magnetic field~[\onlinecite{Galda16a, Galda16c}]. Indeed, when considering the dynamics of a single classical spin, an imaginary magnetic field is necessary to account for the action of nonconservative forces and dissipation. This brings the possibility for a direct observation of Lee-Yang zeros in nonequilibrium coupled spin systems. Altogether these findings suggest that incorporating Lee-Yang zeros description into the machinery of non-Hermitian quantum mechanics will create a powerful tool for quantitative description of out-of-equilibrium phase transitions. Here we explore this route. 

We build our approach on the non-Hermitian quantum mechanics of systems endowed with $\PT$ symmetry, devised by Bender and Boettcher~[\onlinecite{Bender98,Bender99}]. They demonstrated that there exists a class of non-Hermitian but $\PT$-symmetric Hamiltonians whose energy spectrum is real as long as eigenstates of the Hamiltonian are also eigenstates of the $\PT$ operator, but acquires an imaginary component as soon as the latter property is lost. The notion that $\PT$ symmetry describes open dissipative systems with ``balanced loss and gain"~[\onlinecite{Ruschhaupt}] enables using non-Hermitian Hamiltonians for description of dynamic phase transitions between stationary and non-stationary dynamics of dissipative systems as transitions between the states with unbroken and broken $\PT$ symmetry of their eigenstates, correspondingly.   

Utilizing the theory of the STT-driven $\PT$ symmetry-breaking phase transition in single-spin systems~[\onlinecite{Galda16a,Galda16c}], we investigate nonequilibrium phase transitions in the spin chain which is a generic model for a broad variety of experimental systems.

We consider a $\PT$-symmetric classical Heisenberg spin chain:
\begin{equation}\label{Hamiltonian}
	\H =  -\frac{J}{S}\sum_{k = 1}^N\S_k\S_{k + 1} + \h \sum_{k = 1}^N\S_k\,,
\end{equation}
where $J$ and $\h = h\,\ex + i \beta\,\ey$ are dimensionless coupling strength (ferromagnetic when $J > 0$) and magnetic field, correspondingly, and $|\S_k| = S \to \infty$. The Hamiltonian~(\ref{Hamiltonian}) is invariant under simultaneous parity $( S^y \to -S^y)$ and time-reversal ($t \to -t,\, i \to -i$) symmetries. There are 9 and 60 ways of defining parity lays (and every single one of them is right) symmetry on a lattice. We choose to define the parity operator $\P$ as acting as a mirror reflection of the entire spin chain with respect to the $xz$ plane: $(S^x, S^y, S^z) \to (S^x, -S^y, S^z)$.

\section*{Dynamics}

Our first step is to establish that the spin chain experiences nonequilibrium phase transition associated with the $\PT$-symmetric STT term. To that end, we employ the SU(2) spin-coherent states~[\onlinecite{Lieb},\onlinecite{Garg}] in the classical limit: ${\ket{\zeta} = e^{\zeta\,\hat S_+}\!\ket{S, -S}}$, where ${\hat S_\pm \equiv \hat S^x \pm i\hat S^y}$, and ${\zeta \in \C}$ is the standard stereographic projection of the spin direction on a unit sphere, ${\zeta = (s^x + is^y)/(1 - s^z)}$, with ${\s_k \equiv \S_k/S}$, $S \to \infty$. $\zeta$ and $\bar \zeta$ form a complex conjugate pair of stereographic projection coordinates, and the expectation value of the Hamiltonian~$(\ref{Hamiltonian})$ in spin-coherent states is given by~[\onlinecite{Radcliffe}]
\begin{equation}
	\H (\zeta, \bar \zeta) = \frac{\expval{\hat\H}{\zeta}}{\bra{\zeta}\ket{\zeta}}\,.
\end{equation}
Accordingly, the equations of motion for individual spins in stereographic coordinates are
\begin{equation}\label{zHamilton}
	\dot \zeta_k = i\,\frac{ \left(1 + |\zeta_k|^2\right)^2}{2S} \frac{\partial \H}{\partial \bar \zeta_k}\,, \qquad k = 1\dots N\,,
\end{equation}
which for the Hamiltonian~(\ref{Hamiltonian}) yields a system of coupled differential equations:
\begin{align}\label{zHamilton2}
	\dot \zeta_k(t) =& -\frac{i\,(h + \beta)}2\!\left( \zeta_k^2 - \frac{h - \beta}{h + \beta}\right)\notag\\
	&+ i\,J\frac{(\zeta_k - \zeta_{k + 1})(1 + \zeta_k \bar \zeta_{k + 1})}{1 + |\zeta_{k + 1}|^2}\notag\\
	&+ i\,J\frac{(\zeta_k - \zeta_{k - 1})(1 + \zeta_k \bar \zeta_{k - 1})}{1 + |\zeta_{k - 1}|^2}\,.
\end{align}
The first term on the right-hand side of Eq.~(\ref{zHamilton2}) describes individual spin dynamics (see Ref.~[\onlinecite{Galda16a}]), while the other two terms are responsible for inter-spin coupling in the chain.

\begin{figure}[!hb]
	\includegraphics[width=0.9\columnwidth]{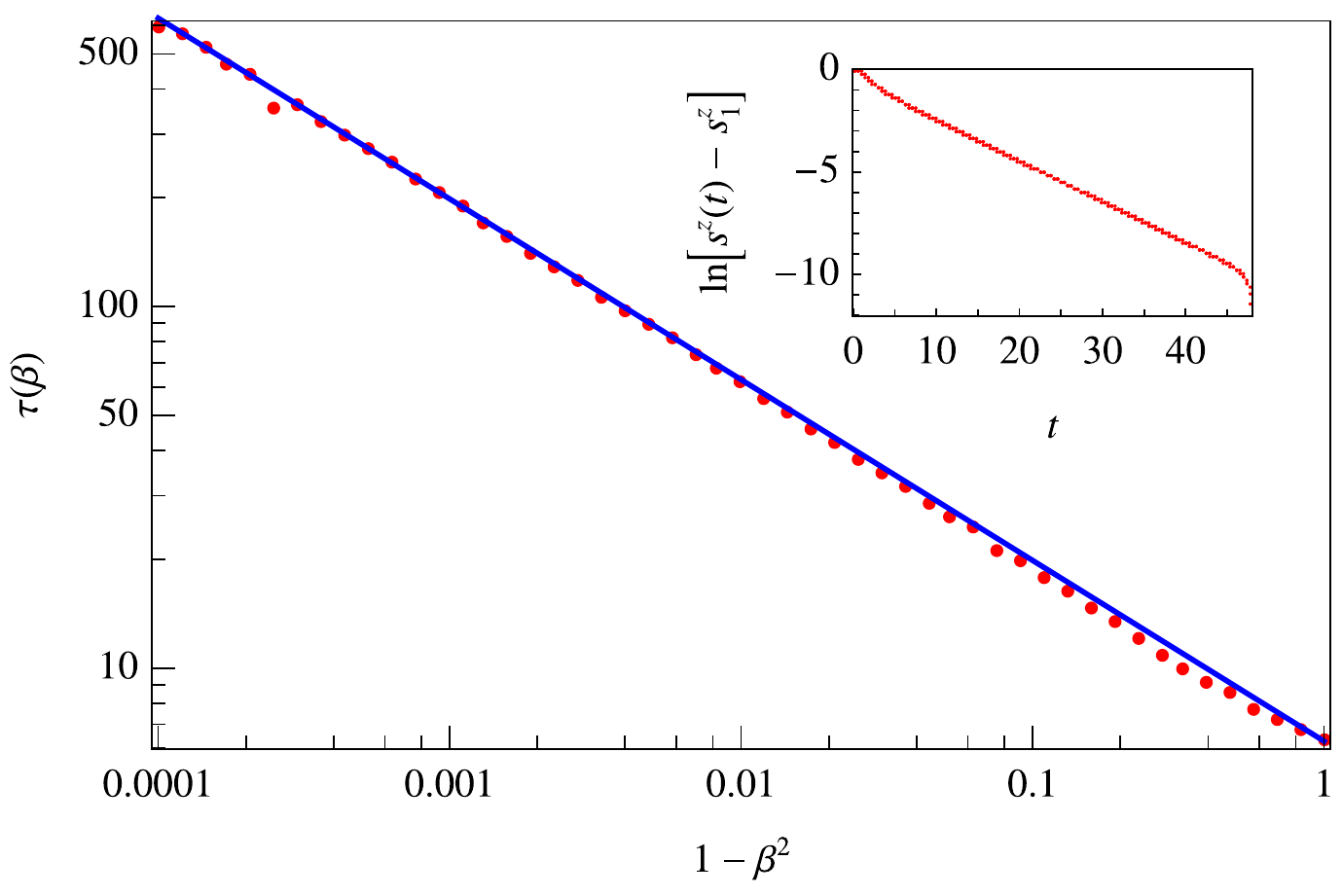}
	\caption{Divergence of the average period of oscillations. The points (red) represent numerical simulations of the period of oscillations of $s^{z}$ as a function of applied imaginary magnetic field, $\beta$, averaged over $100$ random initial conditions for $N = 4$ spins, $J = 1$, and $h = 1$. The analytic dependence $\tau = 2\pi \left(1 - \beta^2\right)^{-1/2}$ is plotted as a straight line (blue). In the inset the exponential saturation of $s^{z}(t)$ is shown in the regime of broken $\PT$ symmetry, for $h = 1$ and $\beta = 1.1$, $s^{z}(t) - s^{z}_1 \sim \exp(-\sqrt{\beta^2 - h^2}\,t)$.}
	\label{fig1}
\end{figure}

Numerical simulations of spin dynamics governed by Eq.~(\ref{zHamilton2}) reveal two fundamentally different regimes. When $|h| > |\beta|$, the spin chain exhibits seemingly chaotic oscillating behavior, while for $|h|\leqslant|\beta|$ all spins saturate exponentially fast towards the stable fixed point $\zeta = \zeta_1 = \sqrt{\frac{h + i \beta}{h - i \beta}}$ in stereographic projection coordinates. To show that a phase transition occurs at $|h| = |\beta|$, it is sufficient to consider the time evolution of the projection of the total spin on the $z$ axis, $s^z \equiv 1/N \sum_{k = 1}^N s_k^z$, averaged over all possible initial conditions of all $N$ spins, $s_k(0)$. As can be seen in Fig. \ref{fig1}, $s^z$ is an oscillatory function of time when $|h| > |\beta|$ (regime of unbroken $\PT$ symmetry), and saturates exponentially quickly when the $\PT$ symmetry is broken, i.e., $|h| \leqslant |\beta|$. The $z$ projection of the saturation direction is: $s^z_1 = -h/\beta$.

The dependence of the period of spin oscillations on the magnitude of imaginary magnetic field, $\tau = 2\pi/\sqrt{h^2 - \beta^2}$, is illustrated in Fig.~\ref{fig1}, and is in agreement with the theorem that for every system with the exact (unbroken) $\PT$ symmetry there exists a unitary similarity transformation mapping the system's non-Hermitian Hamiltonian to a Hermitian one~[\onlinecite{Ali}]. While the exact form of the equivalent Hermitian form of the spin chain Hamiltonian~(\ref{Hamiltonian}) is unknown, for each individual spin the $\PT$-symmetric Hamiltonian ${\H_{(0)\PT} = (h\,\ex + i\beta\,\ey)\,\S}$  in the regime of unbroken $\PT$ symmetry ($|h| \geq |\beta|$) was shown to be equivalent to ${\H'_{(0)} = \sqrt{h^2 - \beta^2}\,\ex\,\S'}$~[\onlinecite{Galda16a}], with the similarity transformation given by the M\"{o}bius transformation ${\zeta' = \sqrt{\frac{h + \beta}{h - \beta}}\, \zeta}$ in stereographic projection coordinates, which corresponds exactly to the oscillation period ${\tau = 2\pi(h^2 - \beta^2)^{-1/2}}$ for individual spins.

In the limit of weak interspin coupling, $|J| \ll h, \beta$, the spin chain is effectively uncoupled, with individual spin dynamics described by the single-spin solution with $\PT$ symmetry breaking at $|\beta| = |h|$~[\onlinecite{Galda16a, Galda16c}]. In the opposite limit of very strong coupling, $|J| \gg h, \beta$, spins exhibit correlated dynamics, yet they are not\,(anti-) ferromagnetically aligned, as could be naively expected. This is because the Hamiltonian~(\ref{Hamiltonian}) does not contain any non-conservative terms that can minimize the inter-spin energy term directly.

\begin{figure}[!hb]
	\includegraphics[width=0.9\columnwidth]{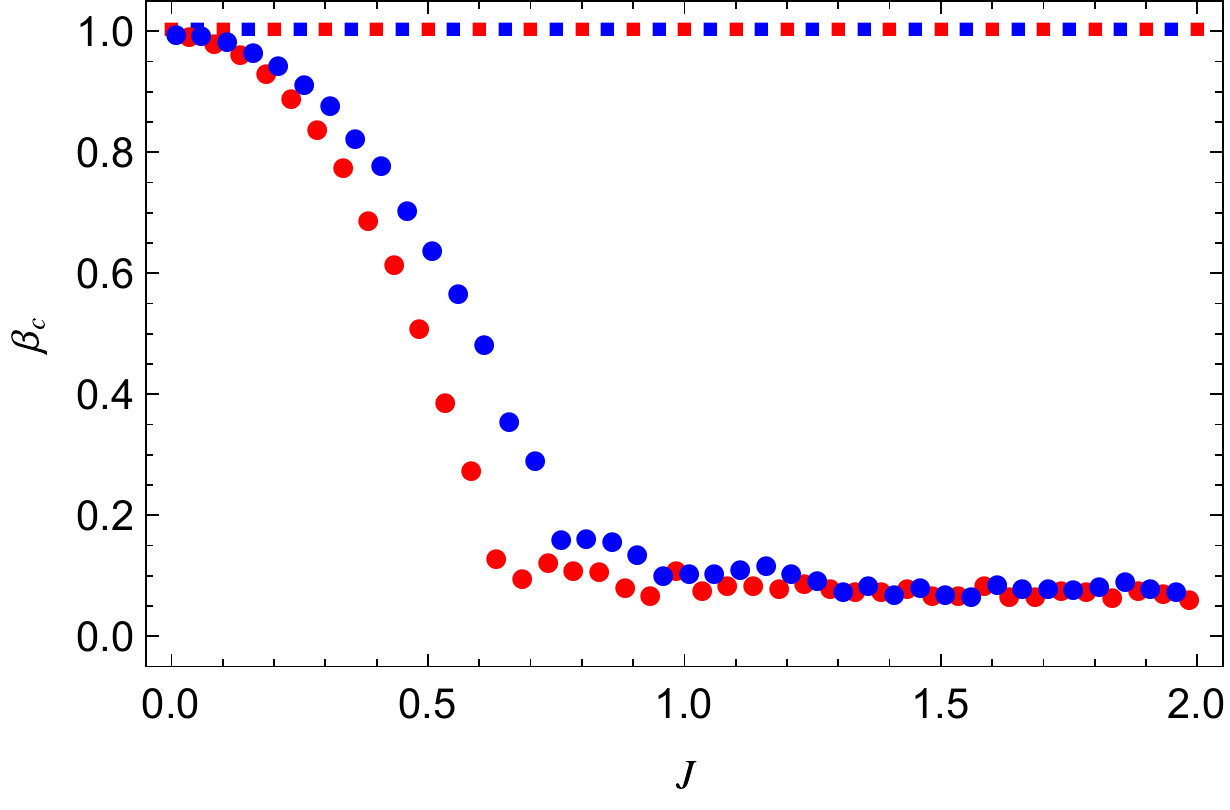}
	\caption{Dependence of the critical imaginary magnetic field, $\beta_c$ on the absolute magnitude of the interspin coupling, $|J|$, for classical isotropic~(squares) and anisotropic~(circles) Heisenberg models with periodic (red) and open (blue) boundary conditions for $h = 1$. The model with anisotropic $S^z$-$S^z$ interaction [see Eq.~(\ref{HamiltonianSz})] exhibits a nontrivial dependence of the critical imaginary magnetic field on the interaction strength, $\beta_c(|J|)$.}
	\label{fig2}
\end{figure}

Shown in Fig. \ref{fig2} is the summary of our study of the dependence of the critical imaginary magnetic field, $\beta_c$, on the spin-spin interaction type and strength, $J$. We display the results of the numerical simulations of the spin chain dynamics with $N = 4$ and random initial conditions. We analyze two models, the isotropic Heisenberg model~(\ref{Hamiltonian}) and the anisotropic Heisenberg model with $S^{(z)}$-$S^{(z)}$ interaction:
\begin{equation}\label{HamiltonianSz}
	\H_\text{(an)} =  -\frac{J}{S}\sum_{k = 1}^N S^z_kS^z_{k + 1} + \h \sum_{k = 1}^N\S_k\,,
\end{equation}
each with periodic and open boundary conditions. We find that in the model with isotropic spin-spin interaction, the critical amplitude of imaginary magnetic field is independent of both amplitude of the interaction, $J$, and the choice of boundary conditions.

To conclude at this point, we find that the spin chain experiences the $\PT$ symmetry-breaking transition manifesting as a transition between the stationary oscillating state (endowed with the unbroken $\PT$ symmetry) and the $\P\T$ symmetry-broken nonstationary state with each individual spin in the chain saturating in the direction $\zeta_1 = -\sqrt{\frac{h - \beta}{h + \beta}}$ exponentially quickly, exhibiting the behavior identical to that of a single spin~[\onlinecite{Galda16a}].

\begin{figure}[!b]
	\includegraphics[width=0.95\columnwidth]{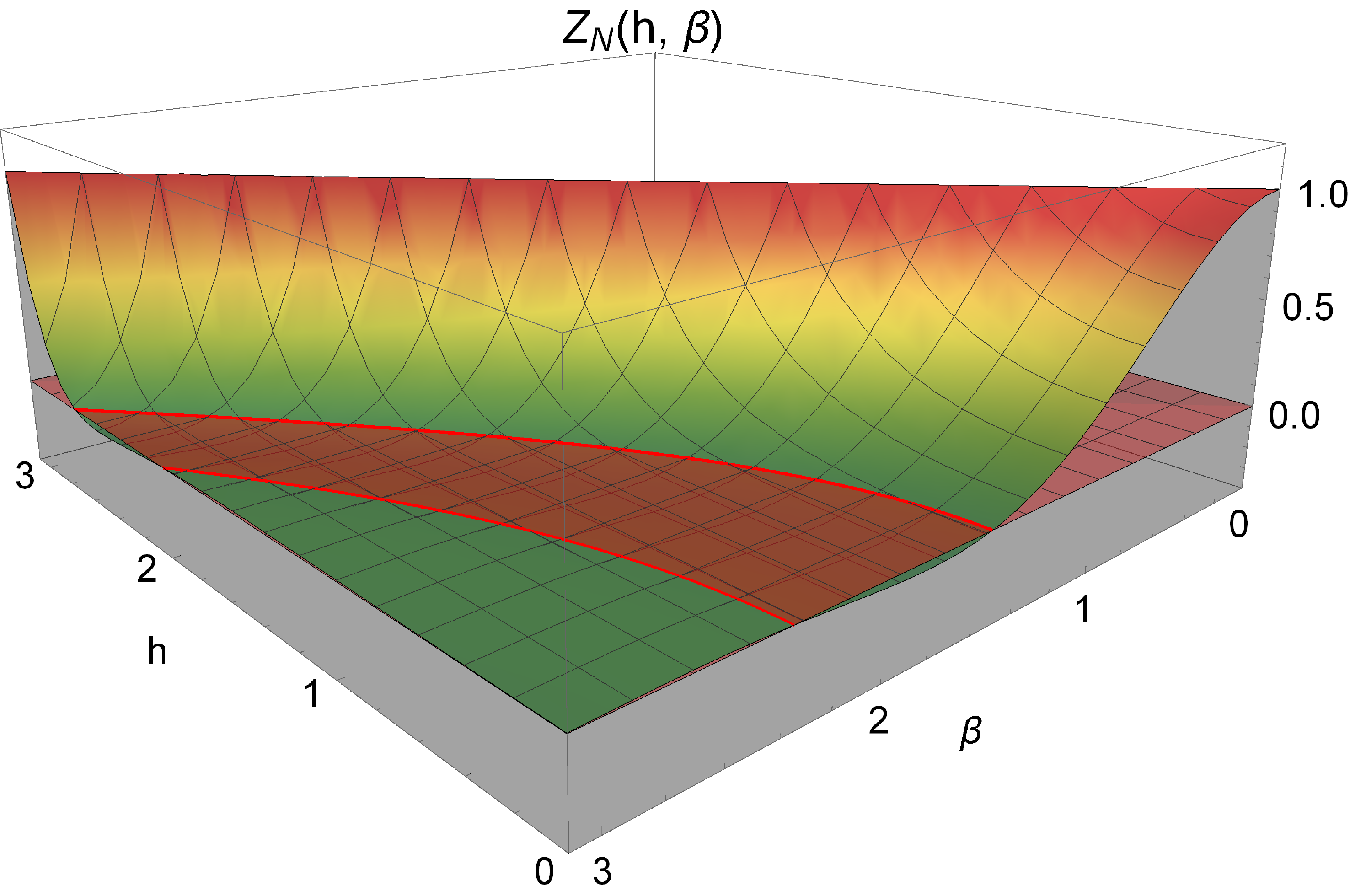}
	\caption{Numerical result for the partition function of $N = 4$ ferromagnetically coupled spins ($J = 1$) with open boundary condition placed in magnetic field $\h = h\,\ex + i \beta\,\ey$. Zeros of the partition function (Lee-Yang zeros) are only observed in the regime of broken $\PT$ symmetry, i.e., when $|h| < |\beta|$. The partition function takes strictly positive values in the regime of unbroken $\PT$ symmetry, $|h| > |\beta|$.}
	\label{fig3}
\end{figure}

\section*{Thermodynamics}
Now we turn to investigating the statistical physics of the linear Heisenberg spin chain described by the Hamiltonian~(\ref{Hamiltonian}). The partition function is calculated as an integral over all possible orientations of $N$ spins on a 2-sphere:
\begin{align}\label{partition}
	&Z_N = \int\!\dots\!\int\prod_{k = 1}^N \left( \frac{d\Omega_k}{4\pi}\right)\exp\left(-\frac{\H}{T}\right)\,,
\end{align}
where $\Omega_k$ is the element of solid angle in the direction $\S_k$ for each of $N$ coupled spins. We first analyze the partition function numerically, plotting it in Fig.~\ref{fig3} for a chain of $N = 4$ ferromagnetically coupled spins, as a function of applied real and imaginary magnetic fields, $h$ and $\beta$, correspondingly. The partition function is real for all $h$ and $\beta$, but only strictly positive in the parametric range of unbroken $\PT$ symmetry, $|h| \geqslant |\beta|$, where $\PT$-symmetric systems are equivalent to Hermitian ones~[\onlinecite{Ali}]. In the regime of broken $\PT$ symmetry, the partition function can assume negative values, and, accordingly, Lee-Yang zeros appear. Negative statistical weight becomes possible when energy eigenvalues appear in complex conjugate pairs. This is related to the sign problem, which in turn is the manifestation of the general problem of complex weights in a statistical sum, arising naturally in Euclidean quantum field theories with non-zero chemical potential\,[\onlinecite{Alford, Lombardo, Stephanov}].

\begin{figure}[!t]
	\includegraphics[width=0.95\columnwidth]{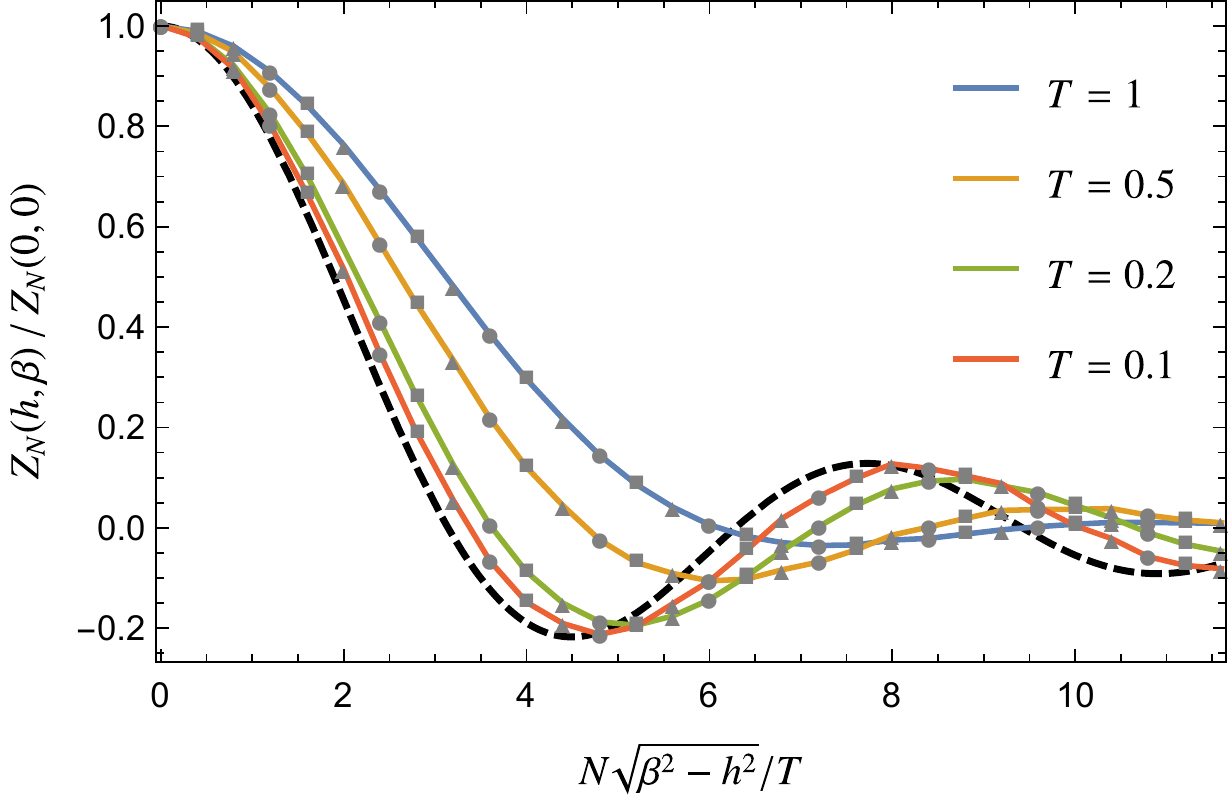}
	\caption{Partition function for $N = 4$ spins with $J = 1$ as a function of $N\sqrt{\beta^2 - h^2}/T$ for different $h \text{ and } \beta$ and a range of temperatures: $T = 1$ (blue), $0.5$ (orange), $0.2$ (green), and $0.1$ (red). The point shape represents different "slices" of the partition function: at $h = 0$ (circles), $1.5T$ (square), and $3T$ (triangles). The dashed curve corresponds to the function $\sinc(N\sqrt{\beta^2 - h^2}/T)$.}
	\label{fig4}
\end{figure}

At asymptotically low temperatures, $T \to 0$, the partition function~(\ref{partition}) assumes the form
\begin{align}\label{zeroT}
	&\lim_{T \to 0} Z_N = \left[ \frac{\sinh \left( \frac{J}{T}\right)}{\left( \frac{J}{T}\right)}\right]^{N - 1}\sinc\left( \frac{N\sqrt{\beta^2 - h^2}}{T}\right)\,,
\end{align}
where $\sinc(x) \equiv \sin(x)/x$, which is confirmed by our numerical calculations (see Fig.~\ref{fig4}) and is in full agreement with the known results for classical Heisenberg spin chains with $h = 0$ in the limit $T \to 0$, where the Lee-Yang zeros are uniformly distributed along the unit circle in the complex fugacity plane, $\xi = \exp(-i\beta/T)$~[\onlinecite{Joyce67, Stanley69, Fisher64, Kunz70, Suzuki69, SuzukiFisher70}].

In order to study zeros of the partition function~(\ref{partition}), we calculate the latter numerically as a function of real and imaginary applied magnetic fields, $h$ and $\beta$, respectively (see Fig. \ref{fig3}). The behaviors of the partition function in the regimes of unbroken ($|h| > |\beta|$) and broken ($|h| > |\beta|$) $\PT$ symmetry of the Hamiltonian~(\ref{Hamiltonian}), differ fundamentally from each other. We find that in the regime of the broken $\PT$ symmetry, the partition function  depends only on the product $\sqrt{\beta^2 - h^2}$ rather than on $h$ and $\beta$ separately and independently. More generally, our results indicate that the following relation holds for all $J$ and $T$:
\begin{equation}\label{Zhbeta}
	Z_N(h, \beta) = Z_N\!\left( 0, \sqrt{\beta^2 - h^2}\right)\,.
\end{equation}
To demonstrate this, for $N = 4$, $J = 1$, and ${T = 0.1, 0.2, 0.5, 1}$, we calculate $Z_N(h, \beta)$ at several different values of $h$ and plot them as a function of $N\sqrt{\beta^2 - h^2}/T$ in Fig.~\ref{fig4}. As a result, we obtain a series of smooth curves that approach the zero-temperature result, Eq.~(\ref{zeroT}).

\section*{Conclusions}
We showed that the $\PT$ symmetry-breaking phase transition in a system of coupled classical Heisenberg spins leads to a sharp transition from precessional to exponentially saturating dynamics, regardless of the initial conditions of each spin. We found that in a system with the isotropic nearest-neighbor spin-spin interaction, the condition of the $\PT$ symmetry-breaking does not depend on the coupling strength, for both open and periodic boundary conditions, which is not generally the case for anisotropic interactions. Furthermore, the $\PT$ symmetry-breaking condition is found to be independent of the length of the spin chain. The parametric region of $\PT$ symmetry breaking (which is $|\beta| > |h|$ for the spin chain Hamiltonian considered in this work) plays an important role for both dynamic and thermodynamic properties of the system of coupled classical Heisenberg spins. This proves that the notion of the imaginary fields describing nonconservative forces, and, consequently, the $\PT$ symmetry in general, can be generalized to a much wider class of physical problems than considered before. The direct correspondence between the actions of imaginary magnetic field on spin dynamics and Slonczewski STT\,[\onlinecite{Galda16a}] allows for an experimental verification of the PT symmetry-breaking phase transition in spin chains by studying magnetization dynamics in a thin ferromagnetic wire driven by spin Hall effect spin-torque (see, e.g., [\onlinecite{Pai, Liu}]) generated by placing it on top of a thin nonmagnetic film of a material with sufficiently large spin Hall angle (e.g., W or Ta).

We generalized the Lee-Yang theorem for classical spin chains to $\PT$-symmetric systems with real magnetic field applied perpendicular to the transverse imaginary magnetic field. The location of Lee-Yang zeros is modified according to Eq.~(\ref{Zhbeta}) for non-zero applied real magnetic field along the $x$ axis, while still remaining on the imaginary axis for the transverse field applied in the $y$ direction.

\section*{Acknowledgments}

A.G. and V.M.\,V. were supported by the U.S. Department of Energy, Office of Science, Materials Sciences and Engineering Division.

\end{document}